\documentstyle[12pt,twoside,fleqn,dina4p]{article}
\def\be{ \begin{equation}}
\def\ee{ \end{equation}}
\def\ba{ \begin{eqnarray}}
\def\ea{ \end{eqnarray}}
\def\F{{\cal F}}
\def\nn{\nonumber}
\newcommand{\ignore}[2]{#2}
\def\rellow#1#2{\mathrel{\mathop{\kern 0pt #1}\limits_{#2}}}
\def\o#1{\rellow{\circ}{#1}}

\def\circY{\rellow{\circ}{Y}}

\def\Mor{Mor}
\def\arrow{arrow}
\def\Arrow{Arrow}

\def\a{\alpha}

\def\A{A}
\def\O{{\Omega}}
\newtheorem{theo}{Representation theorem}

\title{Gauge theory of things alive and universal dynamics
        \thanks{Supported by Deutsche Forschungsgemeinschaft
and German Israel Foundation. Submitted for publication}
}
\author{Gerhard Mack \\
   II. Institut f\"ur Theoretische Physik, Universit\"at Hamburg
}
\begin{document}
\maketitle
\begin{abstract}
Positing complex adaptive systems made of agents with relations
between them that can be composed, it follows that they can be
described by gauge theories similar to  elementary
particle theory and  general relativity. By definition, a universal
dynamics is able to determine the time development of any such system
without need for further specification. The possibilities are limited,
but one of them - reproduction fork dynamics - describes DNA
replication and is the basis of biological life on earth. It is a
universal copy machine and a renormalization group fixed point.
A universal equation of motion in continuous time is also presented.
\ignore{
The laws of mechanics were deduced from the motion of planets, but they
are also valid on earth. The laws of gauge theory werde explored in
the realm of elementary particles, but they are als valid for
living things.}{}
\end{abstract}
 \vspace*{-12.5cm}
\hspace*{10.7cm}
{\large \tt DESY 94-184} \\
\hspace*{10.7cm}
{\large \tt October 1994}
                        \\[12cm]
%
All known interactions between fundamental particles are described
by gauge theories, as we know, including Einsteins theory of gravity,
general relativity. Here we argue that the scope of gauge theory
is much wider, including
\ignore{
\begin{itemize}
}{
\begin{list}{$\bullet$}
{
\parsep0ex
\itemsep0ex
}
}
\item
physical systems
\item
biological organisms
\item
Organizations of human society
\item
spiritual edifices:
Minsky's "society of mind"\cite{Minsky},
languages,
evolutionary algorithms, etc.
\ignore{\end{itemize}}{\end{list}}

{\it Example:\/} Gauge theory of swimming
of micro-organisms \cite{Wilczek}.

Gauge theory can describe
{\it
complex adaptive systems}, i.e. anything  alive in the widest sense,
especially
            {\it autopoietic systems \/} which "make themselves"
in an approximately {\it autonomous\/} fashion
\cite{Maturana}.            All these systems consist of
{\it agents and their relations.} Both evolve in time. Agents
organize themselves into larger structures as a consequence of the
relations between them which determine their interaction.

I will give the argument for gauge theory, present examples for various
of its aspects, and add remarks on important ramifications.

An expanded discussion of the contents of the first two sections is
found in \cite{Mack}.
\section{Structure: What is a thing?}
The state of a system at time $t$ may be considered as a category $K$.
In this way structure can be described.

{\it Agents\/} become objects $X$ of a category,

{\it Relations\/} become arrows $f: X \mapsto Y $ of a category
\\
The {\it basic postulates} of mathematical category theory are

1.
Identity arrows $\iota_X : X \mapsto X  $ exist,

2. Arrows can be composed,
$$ f:X\mapsto Y,g: Y\mapsto Z  \ \  \mbox{defines} \ \
 g\circ f: X \mapsto Z.$$
Composition is associative;
$\iota_Y \circ f =f= f\circ \iota_X .$
\begin{figure}
\begin{center}
\ignore{
\begin{picture}(5,6)(0,0)
}{
\unitlength1.0cm
\begin{picture}(6,1.5)(0,0.7)
\unitlength0.035cm
}
%
%
%
\put(10,80){\mbox{Object:}}
\put(55,77){\frame{\makebox(20,10){}}}
\put(100,80){\mbox{Arrow:}}
\put(140,83){\vector(1,0){22}}
%
%
\ignore{
\multiput(10,10)(22,0){7}{\frame{\makebox(20,10){}}}
}{}
\multiput(10,34)(22,0){7}{\frame{\makebox(20,10){}}}
\ignore{
\multiput(10,58)(22,0){7}{\frame{\makebox(20,10){}}}
}{}
\multiput(20,22)(22,0){6}{\frame{\makebox(20,10){}}}
\multiput(20,46)(22,0){6}{\frame{\makebox(20,10){}}}


%
%
\put(86,39){\circle*{4}}
\put(86,39){\vector(1,0){22}}
\put(86,39){\vector(-1,0){22}}
\put(86,39){\vector(1,1){11}}
\put(86,39){\vector(1,-1){11}}
\put(86,39){\vector(-1,1){11}}
\put(86,39){\vector(-1,-1){11}}
\end{picture} \end{center}
\caption{The structure of a brick wall determines a category}
\end{figure}
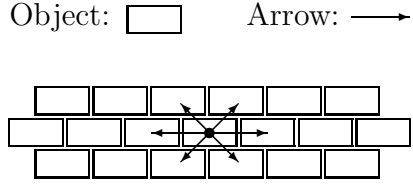
%

{\it Example: brick wall.\/} Its objects are the bricks and the
fundamental arrows specify the translation of a brick to the position
of a nearest neighbour. These arrows specify the structure of the wall.
They can be composed to yield translations to other bricks' positions.

{\it Composability of relations} is our central postulate.
This postulate {\it leads
to gauge theory.}

{\it Examples:}
friend of a friend,
husband of a sister = brother-in-law.

In a {\it *-category}
 a relation $f$ of $X$ to $Y$ defines a possible
 relation $f^{\ast}$ of $Y$ to $X$ (adjoint arrow).

An arrow $f$ is {\it invertible} if an
arrow $f^{-1}$ exists such that
$f\circ  f^{-1}=\iota_Y \ , \ f^{-1}\circ f =\iota_X .$

{\it Functors\/}
         $\F: K\mapsto K^{\prime}$ are maps of categories which preserve
identity and composition law. We call $K^{\prime}$ a representation of
$K$ if $\F$ is onto.

The following special features will be important.

{\bf Locality:}
        Certain arrows are declared as fundamental, and called
{\it links\/}, all others are composed
from them. Dynamics will be such that an agent can be influenced by
another one  if they are related by a link. $\iota_X$ is fundamental.

{\bf Emergence:\/}
Nonlocal phenomena should emerge from
local relations as in physics.

{\it Example: Cognition.\/}
The salient features of a picture (long smooth curves) can be detected
by local interaction among low level cortical neurons which
represent neighbouring picture elements \cite{Ullmann}.

           From links $b_i:X_{i-1}\mapsto X_i$ one composes

           {\bf Paths:\ \ \ }
                $ b_n\circ ...\circ b_2\circ b_1 = C
: X_0\mapsto X_n\ . $  \\
$K$ is {\it connected\/} if there exists a path connecting $X$ to $Y$
for every pair ($X,Y)$.
Closed paths are called {\it loops\/}.

{\it Example:  Lattice Gauge Theory\/} \cite{WilsonKSCreutz}
furnishes a discretized version of the gauge theory of elementary
particles.
The {\it objects\/} are indexed by
sites $x$ of a lattice. They specify matter fields $\Psi(x)\in \O_x$,
where $\O_x$ are vector spaces of some dimension $n$.
The {\it fundamental relations\/} are
      parallel transporters $U[b]:\O_x \mapsto \O_y$
along links $b =(y,x)$ of a lattice. They are unitary linear maps,
possibly respecting additional invariant structure  such as
antisymmetric products $v_1\wedge ... \wedge v_n$.


{\bf Representation theorem: }{\it
 Every finite category admits a faithful representation as a
 communication network as follows: There are spaces $\O_X$ associated
with objects $X$ and arrows act as maps \ $f : \O_X\mapsto \O_Y , $ \
with $\iota_X=id$.
\/}

The construction of the
    space  $\O_X$ uses the sets of all arrows to and
from $X$. Details
are  given in Appendix A.
We talk of one time.
                    The maps $f$ represent channels, not acts, of
communication. Time development  is considered
later.

The maps $f$ need not be linear.
{\it Apart from this, the setup is as in  lattice
gauge theory.\/}
The lattice may be irregular, but irregular lattices were
considered before,
and standard constructs such as covariant Laplacians carry over to
this case. Nonlinearity introduces features of neural nets
\cite{NeuralNets}. There is a fundamental difference to neural net
theory and to any cybernetic approach, though.
Because of the gauge principle
it is in general not appropriate
                      to characterize  an object  at
some time $t$ by what is true about it.

The representation is not unique. Select for every $X$ an invertible
loop $g_X$ and substitute $g_Y\circ f\circ g_X^{-1}$ for
$f : X\mapsto Y$. This produces another, equivalent
representation. These transformations are called
{\it  gauge transformations\/}.
They will be considered in the next section.
\ignore{
{\it Remarks:\/} 1. The reconstruction uses global properties of the
category. It is therefore significant that the physical models and the
last example are formulated at the level of communication networks;
so are therefore their locality properties.\\
2. The question arises whether inequivalent
irreducible representations exist similar
to superselection sectors in local quantum theory \cite{Haag}.
}{}

{\it Remark:} According to Heidegger \cite{He},
                                     the question "what is a thing?"
"{\it is an old one. Always new about it is only that it must  be
asked again and again."\/} I did not answer it; I merely expained
how to describe a thing. The description is in the sprit of
Wittgenstein's isomorphism theory \cite{tractatus}
(see Appendix C).                                   In the last
section of this paper,
        the question will be tied to the problem of how to choose
block spins in renormalization group theory \cite{WilsonRG}, and to a
basic feature (autonomy) of Maturana and Varela's theory of
autopoietic systems \cite{Maturana}.
\section{Gauge theory}
     We generalize the basic notions of gauge theory to the
general setup.

{\bf Curvature = Field Strength = Frustration} \\ exists
unless there is at most
                one arrow $f$ from $X$ to $Y$ for
every $X,Y$.
States in different $\Omega_X$
cannot be compared (without parallel transport which depends on the path
) if there is frustration. Without frustration, all paths from a given
object $X_0$ to  a given $X_n$ define the same arrow..

{\it Example: Gossip\/} results from
 frustrated communication. When it returns from a loop, the
story is no longer the same. Human communication is not necessarily
rational, i.e. limited to exchange of information in terms which have
the same meaning to everybody. Poetry, political propaganda and all
kinds of seduction exploit this.
Gauge theory avoids a priori rationality
posits.

{\bf
     Invariants:} \ $F(\xi_1 , ... \xi_n), \ \ \xi_i\in \Omega_X$
is an  invariant if $F(C\xi_1 , ... C\xi_n)$ is independent of the
path $C: X\mapsto Y$ for every $Y$. It follows that
$F(C\xi_1,...,C\xi_n)=F(\xi_1,...,\xi_n)$ for all loops $C$, since
$C=\iota_X $ is one of them.

 Invariant functions of loops $s:\O_X\mapsto
\O_X$ are similarly defined. More generally, invariants of $K$ can be
defined as {\it functors\/} $\F$ from the network $K$ or from categories
derived from $K$ {\it to unfrustrated categories\/}; for details see
\cite{Mack}.

{\it Remark:\/}                                     The totality
of such functors we  may call {\it meaning\/} of $K$, and the image of
objects,
{\it concepts.\/} Kant, in contradistinction, proposed to transcend
{\it from\/} concepts {\it to\/} objects
in his critique of pure reason \cite{He}.
It is of interest to consider  extended categories which include
besides $K$ also concepts and their relations.
Besides images of relations in $K$ there are relations
                                             from objects to concepts
through the functors $\F$, and there may be additional relations
between concepts from intertwiners between different $\F$'s.
One  may ask how  such extended categories can
grow from $K$ dynamically.
                          (The intertwining relation reads like this.
$T_{12}: \O_{\F_1(X)}^1\mapsto \O_{\F_2(X)}^2 $ is defined, but possibly
zero, for every pair of concepts $\F_i(X)$ which are related to the same
object $X$. If $f: X\mapsto Y$ is an arrow then
$ T \F_1(f) = \F_2 (f) T $. In group theory such intertwiners are made
of Clebsch Gordan coefficients.)

{\it Example: Money }\cite{Luhmann}.
Money in a money-based economy is a prototypical example of an
invariant, and it serves to give an  invariant meaning to certain
transfers in society, viz. acquisition by purchase.
According to Luhmann \cite{Luhmann}, p.69,
         "{\it the medium {\em [money]} assures that
in the realm of economy, actions have approximately the same meaning
for him who acts as for the observer}."

{\bf Consensus} on the meaning of invariants
can be achieved among all agents $Y$ by synchronization =
{\it path independent} parallel transport to Y.

{\it Example:\/} In general relativity, and more generally in
any {\it curved (pseudo) Riemannian manifold\/}
    $M$ there is no global notion of
parallelism of tangent vectors $\xi \in TM$, i.e. no consensus on the
meaning of the direction of a vector is possible. But the length of a
vector is an invariant.

{\bf Gauge group} $G$=$\{G_X\}$=\{ {\it holonomy groups}\}   \\
$G_X$ contains all invertible arrows $f: X\mapsto X$ (loops).
These gauge transformations  map
$\Omega_X$ into $\Omega_X$ and $f:\O_X \mapsto \O_Y$
                                                  into $g_Y\circ f \circ
g_X^{-1}$. So they are invertible
maps of objects and relations.
They are isomorphisms of representations
of the same category $K$ and leave
invariants unchanged. The invariants are the
                                            observables in lattice gauge
theory.

        We could instead define a gauge group $G^{\prime}\supset G$
as the group of all those invertible transformatons of the spaces
                                                     $\O_X $ which
leave all invariants invariant. The holonomy group has the advantage
that    it is intrinsic to the category.

{\it Remark:\/}
For human minds we call the totality of fundamental loops
{\it consciousness\/}. They make the agent aware of himself (because
they transform output into input), and they can serve
as a short time memory (because dynamics will be such that a fundamental
arrow propagates a signal in one time step).
These features characterize consciousness according to
Crick and Koch. The totality of  all loops is related to
{\it perception.}

{\bf Composite objects\/} {\it
                       with internal structure:} Categories can be
objects of new categories. Conversely,
a connected
{\it unfrustrated} subcategory $K_O\subset K$ may be reinterpreted
as a single new composite object. This defines
a new category $K^{\prime }$ which is a representation of $K$. The
required new composition laws are of the form
 $
   g\circ f \ (\mbox{in } K^{\prime}) \ = g \circ C\circ f \ (\mbox{in }
K) \ , $
$C$ a path in $K_O$.

{\it Example: Gauge covariant block spins\/} in lattice gauge theory
\cite{Balaban}.

\section{Gauge transformations in linguistics}
were described by Quine before the gauge theory
of elementary particles was found \cite{Quine}. He says

{\it "the infinite
              totality of sentences of any given speaker's
language can be so permuted or {\em mapped} onto itself  that

                                                       (a) the
totality of speakers  disposition to verbal behavior
 remains
invariant, and yet

 (b) the mapping is no mere  correlation of
sentences with equivalent sentences, in any plausible sense
of equivalence however loose. Sentences without numbers can diverge
drastically from their respective correlates yet the divergences
can systematically so offset one another that
                                      the overall pattern
of associations of sentences with one another  and with non-verbal
stimulation is preserved".}

The disposition to verbal behavior in the presence of a nonverbal
stimulus is the linguists {\it observable}.
                                           Observables are invariant.
The pattern of associations is preserved - i.e. the map is an
isomorphism.

About the role of language, Quine has this to say:
{\it
"This structure of interconnected sentences is a single connected
fabric including all sciences and indeed everything we ever say about
the world".}
Thus, if gauge transformations are here, they are everywhere.

Quine's connected fabric is  interpreted as a category.
The agents are sentences of the language, and the relations are
interconnections or associations between them.
In the spirit of Wittgenstein \cite{tractatus} we may regard
sentences as translations (functorial images) of the world -
or at least an attempt at that.

                                Translations need not exist. Quine
hints at that by giving an example. A tentative english translation
{\it "all rabbits are men reincarnate"\/} of a native sentence would
violate the rules, because a native would assent (it is imagined)
in the presence of any nonverbal stimulus whatever, while an
Englishman would not.

Next
we turn to time development $t\mapsto K_t$. Sorin Solomon proposed to
call it "drama".
\section{Universal Dynamics}
is local dynamics which is defined for {\it every category}.
A state should contain all
necessary information about its time development in itself, without
need for further extrinsic specification.
Gauge invariance is an automatic consequence.

If such a dynamics operates at any level, it will also operate at
more composite levels if it has renormalization group fix point
properties. An example will be shown.

We consider first dynamics in discrete time.
%
The following moves are defined for every (local *-) category
\ignore{
\begin{enumerate}
}
{
\begin{list}{\arabic{enumi}.}
{\usecounter{enumi}
\topsep0ex
\partopsep0ex
\parsep0ex
\itemsep0ex
}
}
\item
death of an object or arrow,
\item
replication of an object or arrow,
\item
fusion of indistinguishable arrows, or objects (inverse of 2),
\item
restitution of a missing adjoint arrow in a *-category,
\item
declaration as fundamental of a composite arrow.
\ignore{
\end{enumerate}
}
{
\end{list}
}
It is easy to describe the new categories also in formulas, but I
omit them.

Dynamics fixes which moves take place next, given the present state
(1st order dynamics) or the state and the moves in the previous time
interval (2nd order dynamics).

There are two types of replication of an object:

2a) with replication of arrows

2b) dividing  arrows among duplicates.

{\it Example 1: Reproduction fork dynamics}
is a version of 2b combined with 4. It is first order.
                       The forks are
 made of arrows without adjoints. They designate
the objects next to be split, see figure 2. The arrows are divided
like this:
ingoing to one copy, outgoing to the other, the other way round
for arrows in forks.                        The dotted arrows are
optional.
\begin{figure}
\begin{center}
\setlength{\unitlength}{0.001875in}%
\begin{picture}(1474,180)(123,460)
\thicklines
\multiput(1150,620)(0.00000,-16.47059){9}{\line( 0,-1){  8.235}}
\put(1150,480){\vector( 0,-1){0}}
\put(1150,620){\vector( 0, 1){0}}
\put(1310,470){\circle*{14}}
\put(1310,630){\circle*{14}}
\put(1160,640){\vector( 1, 0){140}}
\put(1000,460){\vector( 1, 0){140}}
\put(1150,470){\circle*{14}}
\put(1150,630){\circle*{14}}
\put(1000,640){\vector( 1, 0){140}}
\put(1140,620){\vector(-1, 0){140}}
\put(1140,480){\vector(-1, 0){140}}
\put(1300,620){\vector(-1, 0){140}}
\multiput(990,620)(0.00000,-16.47059){9}{\line( 0,-1){  8.235}}
\put(990,480){\vector( 0,-1){0}}
\put(990,620){\vector( 0, 1){0}}
\put(1580,540){\vector(-1, 0){140}}
\put(1440,560){\vector( 1, 0){140}}
\put(860,540){\makebox(0,0)[b]{\smash{$\Longrightarrow$}}}
\put(990,470){\circle*{14}}
\put(1300,480){\vector(-1, 0){140}}
\put(1160,460){\vector( 1, 0){140}}
\multiput(1310,620)(0.00000,-16.47059){9}{\line( 0,-1){  8.235}}
\put(1310,480){\vector( 0,-1){0}}
\put(1310,620){\vector( 0, 1){0}}
\put(990,630){\circle*{14}}
\put(1590,550){\circle*{14}}
\put(1420,540){\vector(-3,-2){ 96.923}}
\put(130,470){\circle*{14}}
\put(140,640){\vector( 1, 0){140}}
\put(280,620){\vector(-1, 0){140}}
\put(300,620){\vector(3, -2){ 96.923}}
\put(410,550){\circle*{14}}
\put(130,630){\circle*{14}}
\put(290,470){\circle*{14}}
\put(290,630){\circle*{14}}
\put(570,550){\circle*{14}}
\put(730,550){\circle*{14}}
\put(400,540){\vector(-3,-2){ 96.923}}
\put(560,540){\vector(-1, 0){140}}
\put(580,560){\vector( 1, 0){140}}
\put(1430,550){\circle*{14}}
\put(1320,620){\vector(3, -2){ 96.923}}
\put(720,540){\vector(-1, 0){140}}
\put(280,480){\vector(-1, 0){140}}
\put(140,460){\vector( 1, 0){140}}
\multiput(130,620)(0.00000,-16.47059){9}{\line( 0,-1){  8.235}}
\put(130,480){\vector( 0,-1){0}}
\put(130,620){\vector( 0, 1){0}}
\put(420,560){\vector( 1, 0){140}}
\multiput(290,620)(0.00000,-16.47059){9}{\line( 0,-1){  8.235}}
\put(290,480){\vector( 0,-1){0}}
\put(290,620){\vector( 0, 1){0}}
\end{picture}
\end{center}
 \caption{Reproduction fork dynamics}
 \end{figure}
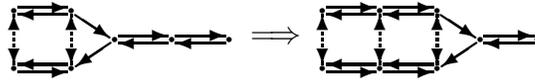
This kind of dynamics is the basis of biological life on earth.
It governs DNA replication during cell division \cite{cell}.
It is a universal copy machine
                            not only for  linear chains or helices, but
                               for {\bf any } structure with
any number of incident arrows on each object and with
substructures of any scale.
                            Therefore it is
a renormalization group fix point \cite{WilsonRG}.
Take any connected *-category without missing
adjoints. Make a single object  replicate. This
creates forks which travel until the whole category has been duplicated,
including any substructure. The dynamics is the same on all scales.

Moreover, the fix point dynamics is insensitive to small perturbations.
Rare random death of arrows or objects will in general
                                            not jeopardize ability
to replicate. Excessive death of arrows leads to random forks, though,
and too many of them  can result in cancerous growth.

A universal copy machine is an optimal prerequisite for evolution.
Programming just a few sterilizations and refertilizations of objects,
elaborate structures can be grown, for instance hypercubic lattices
with periodic boundary conditions, multigrids etc..
                                  A $C^{++}$-class library was written
to implement this \cite{Speh}.
Deviant first replication steps can in some cases induce formation
of  double sheeted covers in place of two copies.

Material properties - embedding of material particles in space -
did not enter here.
              An encyme - topiomerase type II - is known to exist
which relieves DNA from the most serious burden of its material
existence. It enables strings of DNA to pass through each other
\cite{cell}. It is thought to operate by cutting one of the strings,
holding on to the ends while passing them beyond the other string,
and then rejoining them.

{\it Remark 1:\/}
It might be objected that this categorical introduction of replication
begs the question of how the replication of anything works in the
first place. Autocatalytic replication of real molecules is known
\cite{SciAm} and there are simple models.
Suppose the potential energy of unbound "molecule" $a$ in the field of
unbound $b$ has a minimum at binding energy $E$ and a high maximum at
activation energy $A$, while bound or unbound molecules in the field
of bound ones have binding energy $E/2$ and no activation energy. If
any $X=ab$ is present initially, it replicates in an environment
where $a$ and $b$ abound,
$ab + a + b \mapsto  abab \mapsto  ab + ab \ ; $
$E \mapsto 3E/2 \mapsto 2E. $
 \ignore{
$$\begin{array}{rllll}
    E  &    \mapsto & 3E/2 &\mapsto & 2E\ . \nn
\end{array}$$
}{}

{\it Remark2:}
Deterministic replication starting from a single object will yield
indistinguishable clones inside clones.
Stochasticity will alter that, leading to
{\it "identity breaking"\/} in two stages:
from strong identity, where an object is identical only to itself, to
indistinguishability, and from there to distinguishability.
Regarding strong identity as strongest order, each step would
                                                        decrease the
order.

Move 5) is the prototype of  motion.
\begin{figure}
\begin{center}
\setlength{\unitlength}{0.001875in}%
\begin{picture}(1234,137)(243,540)
\thicklines
\put(1460,540){\vector(-1, 0){140}}
\put(480,660){\vector(-2,-3){ 60}}
\put(560,540){\vector(-1, 0){140}}
\put(420,560){\vector( 1, 0){140}}
\put(400,540){\vector(-1, 0){140}}
\put(1320,560){\vector( 1, 0){140}}
\put(860,540){\makebox(0,0)[b]{\smash{$\Longrightarrow$}}}
\put(1240,660){\vector( 2,-3){ 60}}
\multiput(1160,560)(16.47059,0.00000){9}{\line( 1, 0){  8.235}}
\put(1300,560){\vector( 1, 0){0}}
\put(1300,540){\vector(-1, 0){140}}
\put(1140,540){\vector(-1, 0){140}}
\put(1000,560){\vector( 1, 0){140}}
\put(260,560){\vector( 1, 0){140}}
\put(410,550){\circle*{14}}
\put(250,550){\circle*{14}}
\put(490,670){\circle*{14}}
\put(730,550){\circle*{14}}
\put(570,550){\circle*{14}}
\put(1310,550){\circle*{14}}
\put(580,560){\vector( 1, 0){140}}
\put(720,540){\vector(-1, 0){140}}
\put(1150,550){\circle*{14}}
\put(990,550){\circle*{14}}
\put(1230,670){\circle*{14}}
\put(1470,550){\circle*{14}}
\end{picture}
\end{center}
\caption{Move 5 interpreted as motion}
\end{figure}
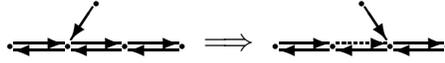

{\it Example2: Quantum mechanical motion of particles in space\/}
can be described by a universal dynamics.
We may picture a category of objects ("space points") linked by
bidirectional arrows plus additional objects ("point particles") linked
to the former by one unidirectional arrow each.
The Schr\"odinger equation for the complex amplitude of a category $K$
reads
$$ i\dot{\Psi}(K)=
    -\Delta \Psi (K)\equiv -\sum_{\mu}[\Psi(\mu K)-\Psi(K)] \ .$$
Summation
is over moves $\mu $
                  of individual particles as in figure 3. For a single
particle of mass $m$
         on a cubic lattice of space points, this is the standard
discretization of the Schr\"odinger equation for free motion. Units
of time $\hbar/2m$ are set to 1.
The "used up" dashed arrow is immediately restituted

{\it Example3: Universal equation of motion in continuous time.\/}
Consider  fundamental arrows
            $f:\O_X\mapsto \O_Y$,
                            including fundamental
                                      loops regarded as attributes of
objects.
Assume spaces $\O_X$  possess tangent spaces $T\O_X$ and maps
                       $f$
                                 are invertible and
have derivatives
$f^{\prime}:T\O_X\mapsto T\O_Y$. Demand
\be   \frac{d}{dt} (\dot{f}\circ f^{-1})  =
    \beta  \dot{f}^{\prime } \circ
   \sum_{g=g_1\circ g_2:X\mapsto Y}
                                     \Bigl\{
( f^{-1\prime}\circ (\dot{g}\circ g^{-1})
  - (g^{-1\prime}\circ \dot{g})\circ f^{-1}
\Bigr\} \label{UNIMOT}\\
\ee
 This is meaningful, i.e. generally covariant (see Appendix B).
It is remarkable that a meaningful equation exists at all.

Summations are over pairs $(g_1,g_2)$ of fundamental arrows so that
$g$ is defined. $\beta \in{\bf R}$ could depend on
whether $f$ is a loop or not.

The equation  is quadratic in velocities
like Einsteins equation for the motion of a massive particle
in a gravitational field,
 $ \ddot{x}^{\mu }
               =-\Gamma^{\mu}_{\ \nu \rho}(x)\dot{x}^{\nu}\dot{x}^{\rho}
\ . $
It retains its form when $f^{-1}$ is substituted for $f$, and induces
similar equations for composite arrows. In this sense it scales.

Note that the expression in $\{ \} $ is only nonzero if there is
nonvanishing frustration, i.e. path dependence, in a generalized sense
as appropriate for initial states $\{ K, \dot{K}\}$.
\ignore{
{\it Scholium:\/}
Let $T\O_X$ the tangent space of $\O_X$ with projector $\pi $.
$\dot{f}:\O_X\mapsto T\O_Y$ with $\pi\circ \dot{f}=f$. Similarly,
$TT\O_X$ is the tangent space of $T\O_X$ with projector $\pi_T$
$h=\ddot{f}$ is a map
\be h: \O_X\mapsto TT\O_Y \ \ \
 \mbox{obeying} \ \ \ \pi_T\circ \ddot{f}=\dot{f} .\label{1}\ee
{\bf R}-linear combinations of such maps are defined and share the
same property.
The derivative
$f^{\prime}:T\O_X\mapsto T\O_X$, obeys the
identity $\pi\circ f^{\prime} = f \circ \pi$, and
$\pi_T\circ \dot{f}^{\prime}=\dot{f}\circ \pi.$
Using these identities
one verifies that the equation of motion is consistent with (\ref{1}).
}{}
\section{Dynamics of composite objects}
should be determined by a
{\bf {\it renormalization group transformation}}
                                                      with dynamically
determined coarse grained variables=blockspins \cite{WilsonRG}.

Grabowski showed recently
how to determine  blockspins dynamically from locality demands
on effective actions in quantum field theory \cite{Grabowski}.
The locality demand amounts to making the composite objects
{\it as autonomous as possible.}

The limits of resolution of  observations will enter into
the specification of the block spins (they fix the scale):
{\it What is a (composite) thing is in the eye of the observer}.
Our mind organizes the world into things which are stable and
                                                capable of an
approximately autonomous existence.

At composite levels, unfrustrated  communication - i.e. communication
by exchange of invariants - may
                            tend to dominate, because of destructive
interference effects for others.

{\it Remark:} This can be seen from random walk expansions.

According to Luhmann \cite{Luhmann},
a system determines its own  medium of [unfrustrated] communication.
"{\it The most important effect of the medium money at the level of
the whole society arises from the fact that the payment acquieses
third parties. Although they are themselves interested in scarce goods
(or could be interested in the future) they are able to look on
{\em [peacefully]} while somebody takes posession of scarce goods,
because he pays for them.}"
Basically, the intrinsically frustrated character of human perception
is partially overcome by assignment of invariant meaning.

%
\section*{Acknowledgments}
I thank H. Joos for pointing out Quine's work, and S. Solomon
and my collaborators M. B\"aker, M. Grabowski,
 M. Griessl, U. Kerres, B. Mikeska, G. Palma, M. Speh and Y. Xylander
 for numerous discussions.

Hospitality at the Santa F\'e Institute and Los Alamos National
Laboratory (R. Gupta),
the Hebrew University (S. Solomon) and the Weizmann Institute
(A. Brandt) is gratefully acknowledged.
\begin{appendix}
\section*{Appendix A: Proof of the main representation theorem}
A slightly more elaborate version of the representation theorem
was proven by the author in \cite{Mack}. For the convenience of the
reader, the statement and proof of this theorem is reproduced here,
and it is shown how the present version, which is closer to
lattice gauge theory, is obtained as a corollary.

We use a slightly more elaborate notation than in the main text.
Given the category $K$
with objects $X,Y,...$, we denote by  $\Mor (Y,X)$ the set of arrows
 $f:X\mapsto Y$ and
  $          g\circY f : X\mapsto Z \  $
for the composite of $f\in \Mor(Y,X)$ with $g\in \Mor(Z,Y)$.
  \begin{theo}\label{DarstSatz}{\em (Representation of a category as a
  communication network)} Every category  $K$ permits a faithful
  representation with the following properties

  To every object $X$ there exists an input space $\A_X $ and an output
  space $\O_X$. The input space contains a distinuished element
   $\emptyset$ ("empty input").
  \Arrow s $f\in \Mor (Y,X), g\in \Mor (Z,Y)$ and
  objects $X$ act as maps
  \ba X:\A_X&\mapsto& \O_X , \\
      \iota_X: \O_X &\mapsto & \A_X\\
      f:\O_X&\mapsto& \A_Y \ea
  with the properties
  \ba X\iota_X=id: \O_X \mapsto \O_X \ &,&
       \iota_X X =id: \A_X \mapsto \A_X \  ,\\
   g\circY f &=& gYf: \O_X \mapsto \A_Z \ . \label{CompMap}\ea
   \end{theo}
 It should be noted that  $\iota_X$ does not act as the identity map
in general in this context.

Given this version of the representation theorem, we restrict attention
to the output spaces $\O_X$ and to maps $\hat{f}= Y\circ f:
\O_X \mapsto \O_Y$. Renaming $\hat{f}$ into $f$
                    we obtain the representation theorem
of the main text.

{\it Remark:}
In some applications, the distinguished empty input is mapped into
a distinguished output which is distinct from all other output. In
this case it may be ignored. This happens in pure
gauge theory without matter fields. In other cases,
$X\emptyset = \Psi_X\in \Omega_X$ is dynamically determined. Matter
fields $\Psi_x$ in lattice gauge theory are an example.
\subsection*{Proof of the
                     representation theorem \ref{DarstSatz}
for categories}
Given a category $K$,
we write  $\Mor (Y,*)$ for the set of all its \arrow s
to $Y$ etc..
              We define
$$ In  (Y) = \Mor  (Y,*) \ \ , \ \ Out  (Y)= \Mor  (*,Y) \ . $$
We write $X=\a     (f)$ if  $f \in \Mor  (Y,X) \subset In  (Y)$,
and correpondingly
  $Z= \omega (f)$ if   $f\in \Mor  (Z,Y)\subset Out  (Y)$. The
output space will be defined as a subspace  $\Omega_Y$ of
                       $\Omega^{virt}_Y$.
$ \Omega^{virt}_Y$ consists of maps
$$ \zeta : Out_Y \mapsto \Mor  (*,*) $$
with the property  $\zeta (f) \in \Mor_K(\omega(f),*)$.

An object $Y$ will act as a map
$$ Y: In  (Y) \mapsto \Omega_Y . $$
according to
$$ Yf(g) = g \o Y f \ \ \ (g \in Out(Y)).  $$
The output space is defined as the image of $Y$, and the input space as
space of equivalence classes (if necessary)
                             of elements of $In_K(Y)$, which $Y$ maps
into the same  $\zeta \in \Omega_Y^{virt}$.
\ba  \Omega_Y&=& IM \  Y \subset \Omega^{virt}_Y \ , \\
     \A_Y &=& In  (Y)/ KER \  Y \ . \label{eqKA} \ea
$Y$ is invertible as a map from $A_Y$ to $\Omega_Y$. Its inverse
is  $\iota_Y$. The empty input $\emptyset \in \A_Y$ is defined as the
equivalence class of  $\iota_Y \in \Mor (Y,Y)\subset In (Y)$.

An \arrow \ $f \in \Mor(Y,X)$ is defined as a map  $\Omega_X \mapsto
\A_Y $ by use use of the map $\iota_X: \Omega_Y
\mapsto \A_Y $, as follows.
\ba f &=& \hat f \o X  \iota_X \ , \\
    \hat f (g) &=& f\o X g \ \  \mbox{ for } g \in \Mor(X,*) \ .
\ea
The last formula defines  $\hat f $ as a map from $In (X)$ to
$In (Y)$. This map passes to equivalence classes (\ref{eqKA})
thereby defining a map                             $\A_X\mapsto A_Y$,
The composition rule
 (\ref{CompMap}) holds.
\section*{Appendix B: Covariance of the universal equation of motion}
To show  that the equation of motion (\ref{UNIMOT}) has a
coordinate independent meaning, we exhibit both sides as elements
of a space which has a coordinate independent meaning.

Let
$E_X=T\O_X$ the tangent space of $\O_X$. It is a fibre bundle over
$\O_X$ with projector $\pi_X $. Then
 $\dot{f}:\O_X\mapsto T\O_Y $ with $\pi_Y\circ\dot{f}=f.$

The  tangent space $TE_X=
 TT\O_X$ is a fibre bundle over  $E_X$ with projector $\pi_{TX}$.
Like the tangent space of every bundle it has a vertical subspace
$VE_X$ which is naturally isomorphic to a factor space
$VE_X\simeq TE_X/    \pi^{\ast}_X T\O_X$ where
$\pi^{\ast }_X: T\O_X\mapsto  TE_X$ is the pullback of $\pi_X$.

The map $f$ induces a map of curves on $\O_X$ into  curves on
$\O_Y$ and thereby a map $f^{\prime}$, called its derivative,
$$f^{\prime}: TO_X\mapsto T\O_Y
      \ \ \mbox{obeying}\ \ \pi_Y\circ f^{\prime}=f\circ \pi_X.$$
Similarly, $\dot{f}$ maps curves in $\O_X$ into curves in $E_Y$. Its
time derivative $\ddot{f}$ is therefore a map from $\O_X$ to $TE_X$
which obeys $\pi_{TY}\ddot{f}=\dot{f}.$ This map passes to the quotient,
also denoted by $\ddot{f}$.Since $\pi_TY$ also passes to the quotient,
$h=\ddot{f}$ is a map
\be h: \O_X\mapsto VE_Y \ \ \
 \mbox{obeying} \ \ \ \pi_{TY} \circ h =\dot{f} .\label{1}\ee
{\bf R}-linear combinations of such maps are defined and share the
same property.

The derivative
$\dot{f}^{\prime}$ of $\dot{f}$ is naturally defined as a map
from $\O_X$ to $TE_Y$. It passes to the quotient, also denoted by
$\dot{f}^{\prime}$, viz.
$$ \dot{f}^{\prime} : \O_X\mapsto VE_Y \ \ \mbox{obeying}\ \
 \pi_{TY}\circ \dot{f}^{\prime}=\dot{f}\circ \pi.$$
Using these identities
one verifies that the equation of motion is consistent with
eq.(\ref{1}) for $h=\ddot{f}$
and is therefore meaningful.

In holonomic coordinates $\{ \eta^{\alpha }, \dot{\eta }^{\alpha }\}$
on $E_Y$,
$$ \ddot{f}(t,\xi ) = \frac{\partial^{2}}{\partial t^2}
f^{\alpha }(t, \xi )\frac{\partial}
    {\partial \dot{\eta}^{\alpha }} \in VE_Y . $$
There is no term proportional to $\frac{\partial}
{\partial \eta^{\alpha }}$ here because
$\pi^{\ast}_YT\O_Y$ consists of such terms.

{\it Remark:} In the notation used here, the chain rule reads
 $ \frac{d}{dt}  g\circ f = \dot{g}\circ f + g^{\prime}\circ \dot{f}.$
\section*{Appendix C: Wittgensteins isomorphy theory \cite{tractatus}}
Wittgensteins propositions  are numbered, therefore every reader may
find them in his own edition of the {\it tractatus\/}
                             in his own language. I give a selection
of those propositions which are particularly relevant here, in german.

{\it Comment 1:} The english translation is problematic; it hides
something that I wish to emphasize. (This only goes to confirm
that translations to other languages may not exist) Wittgenstein
defines {\em Sachverhalt} as a "link ({\em Verbindung}) between
objects
(entities, things)" in proposition 2.01. Ver{\em bind\/}ung comes from
{\em bind}ing. This crucial connection is suppressed by the english
translation of {\em Verbindung} as "combination" instead of link
or bond. {\em Sachverhalt} is translated as "atomic fact". This is an
interpretation due to B. Russell which is motivated by what is said
later in the text. In the same spirit, I propose to compose
general arrows from fundamental ones, called links.

{\it Comment 2:} Wittgensteins remark 5.5303
      on identity is superseded by
quantum mechanics. In his introduction to \cite{tractatus}, Russell
states (p.16/17) [One has] "{\it sought to find such a property\/}
[which must belong to every thing by a logical necessity] {\it in
self identity ...
 {\em [but]} accidental characteristics of the world must,
of course, not be admitted into the structure of logic. Mr.
Wittgenstein accordingly banishes identity ...".}
                  Particles which are identical in the
sense of indistinguishable are very
important in quantum mechanics, and quantum fluctuations introduce
the accidental into the result of measurements.

The issue is  important in the general context of this paper.
Replication of objects, composite or whatever, produces only
indistinguishable copies. Variability comes only from stochasticity.
Mutations such as random death of a component can make formerly
indistinguishable composite objects distinguishable. Sensitive
dependence on initial conditions (chaos) can magnify small random
                                                           fluctuations.
But how do small fluctuations arise in the first place? From
quantum fluctuations. According to the views expressed in this article,
things are in the eye of the observer. The mental organization of the
world into things requires an act of observation.
                                                  This link to
observation brings in quantum fluctuations, because results of
observations on systems in the same quantum state can fluctuate.
\subsection*{Wittgensteins propositions}
\begin{enumerate}
\item
Die Welt ist alles, was der Fall ist.
\begin{enumerate}
\item[1.1]
 Die Welt ist die Gesamtheit der Tatsachen, nicht der Dinge.
\end{enumerate}
\item
Was der Fall ist, die Tatsache, ist das Bestehen von Sachverhalten.
\begin{enumerate}
\begin{enumerate}
\item[2.01]
 Der Sachverhalt ist eine Verbindung von Gegenst\"anden (Sachen,
Dingen).
\begin{enumerate}
\item[2.0121]
  ... so k\"onnen wir uns {\it keinen\/} Gegenstand ausserhalb der
M\"oglichkeit seiner Verbindung mit andern denken.
\end{enumerate}
\item[2.013]
 Jedes Ding ist, gleichsam, in einem Raum m\"oglicher Sachverhalte.
Diesen Raum kann ich mir leer denken, nicht aber das Ding ohne den Raum.
\item[2.021]
 Jede Aussage \"uber Komplexe l\"asst sich in eine Aussage \"uber
deren Bestandteile und in diejenigen S\"atze zerlegen, welche die
Komplexe vollst\"andig beschreiben.
\item[2.024]
      Die Substanz ist das, was unabh\"angig von dem, was der Fall ist,
besteht.
\item[2.032]
 Die Art und Weise, wie die Gegenst\"ande in Sachverhalten
zusammenh\"angen, ist die Struktur der Sachverhalte.
\item[2.033]
Die Form ist die M\"oglichkeit der Struktur.
\\[1mm]
\end{enumerate}
\item[2.1]
    Wir machen uns Bilder der Tatsachen.
\begin{enumerate}
\item[2.12]
     Das Bild ist ein Modell der Wirklichkeit.
\item[2.14]
 Das Bild besteht darin, da\ss \ sich seine Elemente in
bestimmter Art und Weise zueinander verhalten.
\item[2.15]
 Da\ss \ sich die Elemente des Bildes in bestimmter Art und Weise
zueinander verhalten stellt vor, da\ss \ sich die Sachen so zueinander
verhalten.\\
Dieser Zusammenhang der Elemente des Bildes hei\ss e seine Struktur,
und ihre M\"oglichkeit seine Form der Abbildung.
\begin{enumerate}
\item[2.1513]
 Nach dieser Auffassung geh\"ort also zum Bilde auch noch die
abbildende Beziehung, die es zum Bild macht.
\end{enumerate}
\ignore{
2.17 Was das Bild mit der Wirklichkeit gemein haben mu\ss , um sie auf
seine Art und Weise - richtig oder falsch - fabbilden zu k\"onnen
ist seine Form der Abbildung.
\item[2.23]
 Um zu erkennen, ob das Bild wahr oder falsch ist, m\"u ssen wir es
mit der Wirklichkeit vergleichen.
\item[2.225]
      Ein a priori wahres Bild gibt es nicht.
}{}
\end{enumerate}
\end{enumerate}
\item
Das logische Bild der Tatsachen ist der Gedanke
\begin{enumerate}
\item[3.1]
    Im Satz dr\"uckt sich der Gedanke sinnlich wahrnehmbar aus.
\begin{enumerate}
\item[3.12]
     Das Zeichen, durch welches wir den Gedanken ausdr\"ucken, nenne
ich das Satzzeichen. Und der Satz ist das Satz\-zeichen in seiner
projektiven Beziehung zur Welt.
\item[3.13]
 Zum Satz geh\"ort alles, was zur Projektion geh\"ort; aber nicht
das Projezierte ...
\item[3.14]
     Das Satzzeichen besteht darin, da\ss \ sich seine Elemente, die
W\"orter, in ihm auf bestimmte Art und Weise zueinander verhalten.
\begin{enumerate}
\item[3.144]
..\\
(Namen gleichen Punkten, S\"atze Pfeilen ...)
\end{enumerate}
\end{enumerate}
\end{enumerate}
\item
Der Gedanke ist der sinnvolle Satz.
\begin{enumerate}
\begin{enumerate}
\begin{enumerate}
\item[4.001]
      Die Gesamtheit der S\"atze ist die Sprache.
\end{enumerate}
\item[4.01]
     Der Satz ist ein Bild der Wirklichkeit. \\
Der Satz ist ein Modell der Wirklichkeit, so wie wir sie uns denken.
\end{enumerate}
\item[4.1]
    Der Satz stellt das Bestehen oder Nichtbestehen von Sachverhalten
    dar.
\begin{enumerate}
\item[4.21]
 Der einfachste Satz, der Elementarsatz, behauptet das Bestehen
eines Sachverhalts.
\end{enumerate}
\end{enumerate}
\item
Der Satz ist eine Wahrheitsfunktion des Elementarsatzes.
\begin{enumerate}
\begin{enumerate}
\item[5.526 ]
      Man kann die Welt vollst\"andig durch vollkommen verallgemeinerte
S\"atze beschreiben, das hei\ss t also, ohne einen bestimmten Namen
von vornherein einem bestimmten Gegenstand zuzuordnen.
\\
Um dann auf die gew\"ohnliche Ausdrucksweise zu kommen, mu\ss \ man
einfach nach dem Ausdruck  "es gibt ein und nur ein $x$, welches ..."
sagen: Und dies $x$ ist $a$.
\item[5.5303]
       Beil\"aufig gesprochen: Von zwei Dingen zu sagen, sie seien
identisch, ist ein Unsinn, und von einem zu sagen, es sei identisch
mit sich selbst, sagt gar nichts.
\item[6.3432]
       Wir d\"urfen nicht vergessen, da\ss \ die Weltbeschreibung
durch die Mechanik immer die ganz allgemeine ist. Es ist in ihr
z.B. nie von einem {\it bestimmten\/} materiellen Punkte die Rede,
sondern immer nur von {\it irgendwelchen.\/}
\item[6.361 ]
      In der Ausdrucksweise Hertz's k\"onnte man sagen:
Nur gesetzm\"a\ss ige Zusammenh\"ange sind denkbar.
\end{enumerate}
\end{enumerate}
\end{enumerate}
\end{appendix}
%
%

\end{document}